\documentstyle[12pt]{article}  %\def\emptypage{\thispagestyle{empty}~\newpage }
  \oddsidemargin=\evensidemargin
 \let\msk=\medskip

\let\a=\alpha   \let\d=\delta \let\e=\varepsilon
  \let\th=\theta  
    \let\p=\pi \let\r=\rho
   
 \let\ph=\varphi
     
  \let\G=\Gamma 
\def\0{\over }    \def\1{\vec }   \def\2{{1\over2}} \def\3{{\ss}}
\def\4{{1\over4}} \def\5{\bar }   \def\6{\partial } \def\7#1{{#1}\llap{/}}
\def\8#1{{\textstyle{#1}}}        \def\9#1{{\bf {#1}}}
\def\_#1{$\underline{\hbox{#1}}$} \def\^#1{$\overline{\hbox{#1}}$}

\def\<{\langle } \def\>{\rangle }  
 
\def \({\left( } \def \){\right) }

 \let\eq=\equiv    \let\aus=\in
      \let\and=\wedge
     
\def\|#1{{}_{\bigg|_{#1}}}

\def\mao#1{\mathop{\rm #1}\nolimits}      \def\tr{\mao{tr}}

\ifx\Box  \else \fi

\def\pmbf#1{\setbox0=\hbox{${#1}$}   \kern-.025em\copy0\kern-\wd0
      \kern.05em\copy0\kern-\wd0     \kern-.025em\raise.0433em\box0 }
   
  \def\cg{{\cal G}} 
   
  \def\cu{{\cal U}} 
\def\cR{{\cal R}}
\def\inbar{\vrule height1.5ex width.4pt depth0pt} %\font\ZZsf=cmss12
\def\ifundefined#1{\expandafter\ifx\csname#1\endcsname\relax}
\makeatletter \ifundefined{new@mathgroup} {} \else  \input{oldlfont.sty} \fi
\def\ZZ{\relax{\sf Z\kern-.4em \sf Z}}  \def\IR{\relax{\rm I\kern-.18em R}}
\def\IN{\relax{\rm I\kern-.18em N}} \def\IP{\relax{\rm I\kern-.18em P}}
\def\IQ{\relax\,\hbox{$\inbar\kern-.3em{\rm Q}$}}
\def\IC{\hbox{\,$\inbar\kern-.3em{\rm C}$}}
\def\citen#1{\if@filesw \immediate\write \@auxout {\string\citation{#1}}\fi%
\@tempcntb\m@ne \let\@h@ld\relax \def\@citea{}%
\@for \@citeb:=#1\do {\@ifundefined {b@\@citeb}%
    {\@h@ld\@citea\@tempcntb\m@ne{\bf ?}%
    \@warning {Citation `\@citeb ' on page \thepage \space undefined}}%
    {\@tempcnta\@tempcntb \advance\@tempcnta\@ne
    \setbox\z@\hbox\bgroup\ifcat0\csname b@\@citeb \endcsname \relax
       \egroup \@tempcntb\number\csname b@\@citeb \endcsname \relax
       \else \egroup \@tempcntb\m@ne \fi \ifnum\@tempcnta=\@tempcntb
       \ifx\@h@ld\relax \edef \@h@ld{\@citea\csname b@\@citeb\endcsname}%
       \else \edef\@h@ld{\hbox{--}\penalty\@highpenalty
	      \csname b@\@citeb\endcsname}\fi
    \else \@h@ld\@citea\csname b@\@citeb \endcsname \let\@h@ld\relax \fi}%
 \def\@citea{,\penalty\@highpenalty\hskip.13em plus.13em minus.13em}}\@h@ld}
\def\@citex[#1]#2{\@cite{\citen{#2}}{#1}}%
\def\@cite#1#2{\leavevmode\unskip
  \ifnum\lastpenalty=\z@\penalty\@highpenalty\fi% highpenalty before
  \ [{\multiply\@highpenalty 3 #1%              % triple-highpenalties within.
  \if@tempswa,\penalty\@highpenalty\ #2\fi}]}   % and before note.
\makeatother % END OF   "CITE.STY"   (\citen can be used for citation numbers)

\def\beq{\begin{equation}} \def\eeq{\end{equation}} \def\eql#1{\label{#1}\eeq}
\def\bea{\begin{eqnarray}} \def\eea{\end{eqnarray}} 
\def\fnote#1#2{\begingroup\def\thefootnote{#1}\footnote{#2}
	   \addtocounter{footnote}{-1}\endgroup}    

\def\plb#1 #2 {Phys. Lett. {\bf B#1} #2 }
\def\phr#1 #2 {Phys. Rep. {\bf  #1} #2 } 
\def\npb#1 #2 {Nucl. Phys. {\bf B#1} #2 }
\def\aph#1 #2 {Ann. Phys. {\bf #1} #2 }  
\def\jmp#1 #2 {J. Math. Phys. {\bf #1} #2 }
\def\prd#1 #2 {Phys. Rev. {\bf D#1} #2 }
\def\prl#1 #2 {Phys. Rev. Lett. {\bf #1} #2 }
\def\rmp#1 #2 {Rev. Mod. Phys.  {\bf #1} #2 }
\def\zpc#1 #2 {Z. Phys. {\bf #1C} #2 }
\def\cmp#1 #2 {Comm. Math. Phys. {\bf #1} #2 }
\def\mpl#1 #2 {Mod. Phys. Lett. {\bf A#1} #2 }
\def\ijmp#1 #2 {Int. J. Mod. Phys. {\bf A#1} #2 }

\let\3=\ss \catcode`\"=\active \let"=\"

\long\def\del#1\enddel{ }
\ifundefined{emptypage} \let\emptypage=\relax \fi
 \def\[{\left[} \def\]{\right]}

\def\tbf#1:{{\noindent\bf #1:}} \def\new#1\endnew{{\bf #1}}

\begin{document}
\def\tuwien{TUW-94/01} \def\cern{CERN-TH.7165/94} \def\hepth{hep-th/9402114}
\def\pha#1,#2.{\ph_{#1}^{(#2)} }     \def\sig#1{(-1)^{#1}}
\def\detat#1,#2.{\det #1_{|_{#2}}}   \def\ac#1>{|#1\>_{(a,c)}}
\def\^#1{\overline{#1}} \def\A{{\5\a}} \def\X{{\^X}} \def\W{{\^W}}

{\hfill\hepth\vskip-9pt \hfill\cern\vskip-9pt \hfill\tuwien} \vskip 15mm

\centerline{\hss\Large\bf The Mirror Map for Invertible LG Models \hss}

\begin{center} \vskip 8mm
      ~Maximilian KREUZER\fnote{*}{e-mail: kreuzer@dxcern.cern.ch}$\,$
\vskip 5mm
       CERN, Theory Division\\
       CH--1211 Gen\'eve 23, SWITZERLAND
\vskip 3mm               and
\vskip 3mm
       Institut f"ur Theoretische Physik, Technische Universit"at Wien\\
       Wiedner Hauptstra\3e 8--10, A--1040 Wien, AUSTRIA
       \fnote{\#}{Permanent address}

\vfil                        {\bf ABSTRACT}                \end{center}

Calculating the (a,c) ring of the maximal phase orbifold for `invertible'
Landau--Ginzburg models, we show that the Berglund--H"ubsch construction
works for all potentials of the relevant type.
The map that sends a monomial in the original model to a twisted state in the
orbifold representation of the mirror is constructed explicitly.
Via this map, the OP selection rules of the chiral ring exactly correspond to
the twist selection rules for the orbifold.
This shows that we indeed arrive at the correct point in moduli space, and that
the mirror map can be extended to arbitrary orbifolds, including non-abelian
twists and discrete torsion, by modding out the appropriate quantum symmetries.

\vfil\noindent \cern\\ \tuwien\\ February 1994 \msk
\thispagestyle{empty} \newpage  \emptypage
\setcounter{page}{1} \pagestyle{plain}         \baselineskip=14pt

\section{Introduction}

At the level of the effective 4-dimensional field theory of a compactified
heterotic string, mirror symmetry \cite{dix,lvw,gp,cls} -- a redefinition of
the left-moving $U(1)$ charge of the internal $N=2$ superconformal field
theory -- boils down to charge conjugation.
Nevertheless, this transformation has attracted considerable interest,
because it allows for the computation of Yukawa couplings \cite{co}
that are otherwise inaccessible in the Calabi--Yau
compactification scheme. From this geometrical point of view, mirror symmetry,
which now changes the topology of a manifold, is much less obvious. And, in
addition to its computational use, it may serve as an indicator for
non-completeness of classes of compactifications.

Much of what we said about Calabi--Yau manifolds is also true for
Landau--Ginzburg models~\cite{mvw}, which provide sort of a bridge between
the geometrical framework and superconformal field theory~\cite{gvw}.
Here an $N=2$ theory is described by a quasi-homogeneous superpotential,
which, according to non-renormalization theorems, contains the complete
information about the chiral ring~\cite{lvw}.
In order to use such  $N=2$ theories for constructing supersymmetric string
vacua we first have to project to integral (left) charges $J_0$~\cite{v}.
This modding by $j=\exp{2\p iJ_0}$ leads to new states from twisted sectors,
some of which may end up in the (a,c) ring, i.e. the ring of  primary
fields whose left-moving components are anti-chiral and thus have negative
left charge. All of these rings are connected among
each other and with the Ramond ground states by spectral flow~\cite{lvw}.

Of course we can twist the model by more general symmetries of the potential,
and it is only if the resulting orbifold has integral left {\it and} right
charges that we may have a geometrical interpretation, with the chiral ring
corresponding to the cohomology ring of a K"ahler manifold.
The resulting class of (2,2) models lacks mirror symmetry in a significant
way~\cite{nms,aas}, but a large subclass of the potentials,
which we call `invertible', leads to perfectly symmetric spectra.
These potentials are defined by the property that the number of monomials is
equal to the number of fields. Non-degeneracy then implies \cite{cqf}
% \footnote% {See for example \cite{cqf} and references therein.}
that all these monomials are either of Fermat type $X_i^{a_i}$, or of the form
$X_i^{a_i}X_j$. Invertible potentials thus consist of connected closed loops
of `pointers' from $X_i$ to $X_j$, or of chains that terminate with
a Fermat type monomial,
\beq W_{loop}=X_1^{a_1}X_2+\ldots+X_{n-1}^{a_{n-1}}X_n+X_n^{a_n}X_1, \eeq
\beq W_{chain}=X_1^{a_1}X_2+\ldots+X_{n-1}^{a_{n-1}}X_n+X_n^{a_n}.   \eeq
This implies a natural notion of `transposition' by inverting the
direction of the `pointers' while keeping the exponents $a_i$ attached
to the fields $X_i$, or the other way round \cite{bh}.

It is exactly for this class of models that a construction of the mirror as a
particular orbifold of the `transposed' potential has been suggested some time
ago by Berglund and H"ubsch (BH) on the basis of systematic
observation~\cite{bh}. In that paper the correct twist group was found
by demanding that the so called geometrical and quantum symmetries~\cite{q}
be exchanged when one goes to the mirror.
More recently, the BH construction has found an interpretation and a
generalization for Calabi--Yau
hypersurfaces of toric variaties, which are described
by certain families of rational convex cones~\cite{bat}.
% Further support for the BH conjecture comes from a recent paper~\cite{elli},
% in which the matching of the elliptic genera of the proposed mirror pairs is
% checked.
In a recent paper~\cite{elli}, the matching of the elliptic genera of the
proposed mirror pairs has been checked.
{\it Assuming} that the charges of all states that contribute to this
index are anti-symmetric for the orbifold partner, this implies that at
least the charge degeneracies are indeed correct.

A general proof for this construction in the
Landau--Ginzburg framework, and an understanding of the mechanism by
which it works, however, is still missing.
In the present note we try to fill this gap.
Our approach is basically along the lines of the first proof of mirror
symmetry by Greene and Plesser~\cite{gp}, who used the results of
ref.~\cite{gq} on modular invariants of parafermionic theories to obtain a
construction for tensor products of minimal models, i.e.
the case of Fermat type potentials.
All we have to do is to compute the charge degeneracies of the (a,c) ring
for the maximal phase orbifold of an arbitrary connected component of an
invertible potential. To this end we use the formulas for Landau--Ginzburg
orbifolds that were derived by Intriligator and Vafa~\cite{iv}.

A necessary condition for arriving at the mirror of the original $N=2$
theory is that all (c,c) states are projected out.
It is therefore not surprising that the construction works just for the
invertible potentials, which have the minimal number of monomials and thus
a maximal symmetry.
We will find that the states in the (a,c) ring turn out correctly and we will
explicitly construct the map from monomials to twisted states for
the two cases of chains and loops.
The consistency of this map with the original ring structure and the twist
selection rules will then establish that we indeed arrive at the correct point
in moduli space of the correct conformal field theory.

This result immediately implies
that the construction can be extended to arbitrary orbifolds of tensor
products of such models, including non-abelian twists and discrete torsion.%
\footnote{For abelian orbifolds of minimal models this was verified
          in ref. \cite{ade}.}
All we need to do is to mod out the corresponding symmetries of the mirror
partners. In the orbifold representation, the phase part of these symmetries
is realized as a quantum symmetry, which can be implemented by introducing
appropriate discrete torsions with transformations that may act trivially
otherwise
% (the formal order of these transformation is given by the quanta
% of their discrete torsions;
(effectively, this often just undoes part of the orbifolding).
Permutation symmetries are, of course, `geometrical' in both cases.

Assuming that the BH construction gives the mirror at the correct point
in moduli space, it is not only natural to expect that it should also apply
to the (left-right symmetric) original Landau--Ginzburg model, but it even
has to be so. The reason is that, as discussed above, we can undo the
projection to integral charges, and we can do the analogous thing for the
mirror partner, which is an isomorphic conformal field theory by assumption.
But then also the mirror theory has to factorize, and we conclude that
the construction should work for each connected component of the potential
individually.

As a final, technical point, note that we can do the calculation either in
the Ramond or in the Neveu-Schwarz sector. Once we have established that the
charges are anti-symmetric in the proposed mirror model, we can get the other
sector by spectral flow. The Ramond ground states may seem to be the easier
choice, because then we have a single formula for all states and need not
worry
about two different rings. Nevertheless, we prefer the NS sector, because
the ring structure and the selection rules that we expect will help us find
the mirror map more easily. Furthermore, the identification of twisted
states of the maximal orbifold with monomials in a different LG model (with
asymmetric charges) may provide new information about the (a,c) ring
of arbitrary phase orbifolds.

In the next section we will recall some results of ref.~\cite{iv}, which
provide the basis for our computation of the (a,c) rings for $W_{loop}/\cg$
and $W_{chain}/\cg$ in sections 3 and 4, respectively.
In the final section we summarize our results and discuss some implications.

\section{Orbifold setup: the (a,c) ring}

In their semiclassical analysis
% , using the invariance of the index $\sig F$ under modular transformations,
Vafa and Intriligator~\cite{v,iv} derived
all ingredients that are necessary for calculating the charges and
transformation properties of the Ramond ground states and of
their (anti-)chiral relatives for arbitrary Landau--Ginzburg orbifolds.
In this section we briefly recall from their results what we will need in the
following.

As we decided to work in the NS sector, we first need to make sure
that no states survive the projection in the (c,c) ring. Setting the discrete
torsions $\e(g,h)$ and the factor $\sig{K_g}$, which determines the sign of the
group action in the Ramond sector, all equal to 1, it can be shown that
the $h$-twisted vacuum $|h\>_{(c,c)}$ transforms with a phase
$(\detat g,h.)/(\det g)$ under a generator $g$ of the twist group.
Here $\detat g,h.$ is the determinant of $g$ when restricted to the
superfields $X_i$ that are invariant under the twist $h$. These fields
are also the ones that generate the chiral ring
of the  unprojected $h$-twisted sector, with the
vanishing relations derived from their effective Landau--Ginzburg potential.
We thus have to make sure that no monomial in the resulting chiral ring
transforms with a phase that
compensates the above phase factor, thereby forming an invariant state.
Let us anticipate that this is indeed the case, which is a simple
consequence of the results that we will derive below.

The situation in the (a,c) ring is more complicated.
In ref. \cite{v} it has been observed that the
asymmetric spectral flow operator $\cu_{(-1,0)}$ is identical to the field
that introduces a $j$-twist. Therefore
\beq \ac h>=\cu_{(-1,0)}|h'\>_{(c,c)}=\cu_{(-\2,\2)}|h'\>_R ~~~~ \hbox{ with }
         h'= hj^{-1}, \eeq
and the unprojected ring in the sector $h$
is generated by the fields that are invariant under
$h'$. The spectral flow operator $\cu_{(-1,0)}$ shifts the left charge by
$-c/3$, where $c=3\sum_i(1-2q_i)$ is the central charge.
Let $0<\th_i^h<1$ be the phases of the fields $X_i$ that are not invariant
under the (diagonal) action of a symmetry transformation
$h$, i.e. $hX_i=\exp(2\p i \th_i^h) X_i$.
Then the left  and right charges $Q_+$ and $Q_-$
% (i.e. the eigenvalues of $J_0$ and $\5J_0$)
of the vacuum in the $h$-twisted sector of the (a,c) ring are given by
\beq Q_\pm \ac h> = %{J_0\atop \5J_0} =
   \(\mp{c\06}\pm\sum_{tw}(\th_i^{h'}-\2)-\sum_{inv}(\2-q_i)\)\ac h>,\eql{qpm}
where the second and the third term are the contributions of the fields that
are twisted and invariant under $h'$, respectively.
The first term is the charge of
$\cu_{(-1/2,1/2)}$ and thus shifts the result for the Ramond ground states to
the (a,c) ring.
As an example, it is easy to check that the vacuum $\ac 0>=|0\>$ has
vanishing charges: All fields are twisted, and $\th_i^{h'}=1-q_i$.
In the sector twisted by $j$, on the other hand, all fields are invariant
and we find $Q_+=-{c\03}$ and $Q_-=0$.

In the following we will always be interested in the situation that the states
in the (a,c) sector should have anti-symmetric charges. To check this,
we will first have to show that the only monomials surviving the projection
are the ones that have charge $Q_\pm=\sum_{inv}(\2-q_i)$. This will imply
that $J_0=-\5J_0$ in the (a,c) ring.
Using $c/6=(\sum_{tw}+\sum_{inv})(\2-q_i)$, we can simplify
the formula (\ref{qpm}) for the right charge $Q_-$ and obtain
\beq \5J_0\ac h>=\sum_{tw} (1-q_i-\th_i^{h'})\ac h>. \eql{qright}
For the projection to invariant states we need, as our final ingredient,
the action of an arbitrary group element $g$ on the twisted vacuum in the
sector $h$,
\beq g\ac h>=\sig{K_gK_h}\e(g,h)\detat g,h'.\ac h>,\eql{inv}
which was derived in \cite{iv} using the modular invariance of the index
$\tr_R \sig F$ and spectral flow.
As long as we do not consider general orbifolds and the modding of
quantum symmetries, we can set $\e(g,h)=\sig{K_g}=1$.

\section{Loop potentials }

Because of its more symmetric form we first construct the mirror map for
the potential
\beq W_{loop}=\sum_{i=1}^n X_i^{a_i}X_{i+1}, \eeq %~~~~~ X_{i+n}\eq X_i, \eeq
where all indices are defined modulo $n$.
Quasi-homogeneity implies $a_iq_i+q_{i+1}=1$ for the $U(1)$ charges $q_i$ of
the $N=2$ superfields $X_i$
(we restrict the exponents to $a_i>1$ so that $0<q_i<\2$).
For any phase symmetry $\r$, acting as    \beq \r X_j=e^{2\p i\ph_j}X_j, \eeq
the phase $\ph_j$ of the field $X_j$ determines the phase $\ph_{j+1}\eq-a_j
\ph_j$ modulo 1. Therefore all diagonal symmetry groups of the potential are
cyclic. Of course the charges $q_i$ solve these equations, but in general
the corresponding symmetry, generated by $j=\exp(2\p i J_0)$, is only a
subgroup of the maximal phase symmetry.

To obtain a generator $\r_i$ of the maximal group $\cg$ with order $\G=|\cg|$,
we choose the phase to be minimal, i.e. $\pm1/\G$, for some field $X_i$.
It will be useful to choose the sign as $\sig{n}$, because then the phase
of the determinant of $\r_i$ is negative (see below).
In this way we obtain a family of generators $\r_i$ with corresponding phases
$\pha j,i.$ given by
\beq \r_iX_j=e^{2\p i\,\ph^{(i)}_j} X_j, ~~~~~~ \ph^{(i)}_{i+j}={\sig{n-j}
     a_i\ldots a_{i+j-1}\0\G} ~~~ \hbox{for} ~~ 0\le j<n \eeq
(recall that indices are identified modulo $n$). If we set $j=n$ in this
formula we should get a phase that is $\sig n/\G$ modulo 1, thus we obtain
\beq \G=A-\sig n, ~~~~~~~ A=a_1a_2\ldots a_n. \eeq
This is close to, but not exactly the dimension of the chiral ring
\beq |\cR|=\prod {1-q_i\0q_i}=A, \eeq
where we have used $a_i=(1-q_{i+1})/q_i$. We will see later on how the counting
of states work out correctly, in spite of this apparent mismatch.
A convenient basis for the chiral ring is given by the monomials
$\prod X_i^{\a_i}$ with $0\le\a_i<a_i$. This set has the correct number of
elements, and it is easy to see that, using $\6W/\6X_i=0$, all non-vanishing
monomials can be brought into this form.

Since all $\r_i$ generate the same group, they must be powers of one another,
and we find
\beq \r_i=(\r_{i+1})^{-a_i}, ~~~~~~ \pha l,i.+ a_i\pha l,i+1.=-\d_l^i. \eeq
Summing over $l$ we see that the quantities
\beq \5q_i=-\sum_{l=1}^n\pha l,i.={-1\02\p i}\ln\det\r_i \eql{qibar}
satisfy the relations $\5q_i+a_i\5q_{i+1}=1$ and thus they are the weights of
the chiral fields $\X_i$ in the $N=2$ theory with the `transposed' potential
\beq \W_{loop}=\X_1\X_2^{a_1}+\ldots+\X_{n-1}\X_n^{a_{n-1}}+\X_n
     \X_1^{a_n}. \eeq
This is our first hint that, as is well-known~\cite{bh}, the orbifold
$W/\cg$ should be compared to the model described by the transposed potential
$\W$.

It is now easy to see how $j$ is related to the generators $\r_i$. Here we use
the relation
\beq \pha l+1,i. +a_l\pha l,i.=-\d_{l+1}^i. \eeq
Summing over all $i$ we find
\beq q_l=-\sum_{i=1}^n\pha l,i.,~~~~~~ j^{-1}=\r_1\ldots\r_n. \eeq
Without calculation this could have been concluded also from the fact
that the phases $\5\pha i,l.$ of the phase symmetries $\5\r_l$,
which act on the transposed potential, coincide with $\pha l,i.$.

With these formulas it is now easy to construct the mirror map explicitly.
The natural candidate for the image of $\X_i$ is the twisted ground state
$\ac\r_i>$. It is obvious from (\ref{qibar}) that its right charge
(\ref{qright}) coincides with $\5q_i$ up to an integer.
Exact equality then follows from the inequalities $q_j-1\le \pha j,i.\le q_j$,
which will be derived in the next section:
All contribution to the right charge are of the form $1-q_i-\th_i^{h'}=
1-q_j-(1+\pha j,i.-q_i)=-\pha j,i.$, because $\pha i,j.-q_i$
is always between $-1$ and $0$.
Invariance of these states under the group $\cg$ is obvious from
eq.~(\ref{inv}) because all fields are twisted.

{}From the ring structure associated to $\W$ we thus conclude that a monomial
$\prod\X_i^{\A_i}$ should be mapped into a sector twisted by
$h=\prod\r_i^{\A_i}$, where $0\le \A_i<a_{i-1}$.
To see what twists we get in this way, it is sufficient to calculate the phase
of $X_1$, which is given by $\sum\A_i\pha1,i.=n/\G$ for some integer $n$. It
is important to know this number exactly, and not just modulo $\G$.
Inserting the above formulas for $\pha i,l.$ and $q_1$ we see that $n$ takes
all values between $n_-$ and $n_+$ exactly once, where
$n_+=\G q_1=n_-+\G$ if $n$ is even, and
$n_++1=\G q_1=n_-+\G-1$ if $n$ is odd.
If $n$ is even we thus get the twist $h=j$ twice, whereas for $n$ odd this
twist does not occur at all in the image of the mirror map. All
other twists arise exactly once.
% Thus we get all twists $h\neq j$ exactly once, and we get the twist $h=j$
% twice if $n$ is even and this twist does not occur if $n$ is odd.
Since $X_1$ is in no way distinguished, it follows that $q_i-1\le\ph_i\le q_i$,
where the particular representation of a group element $\r$ in terms of
certain powers of the generators $\r_i$ is, of course, crucial.

To see whether this map is one-to-one we now have to look at what happens to
the sector twisted by $j$. Using the generator $\r_1$ of $\cg$
for the projection, the phase $-\5q_1$ of $\det \r_1$ will have to be
compensated by the phase of a monomial $\prod X_i^{a_i}$ if a state in that
sector should survive. Fortunately, we can determine the possible phases
without further calculations if we use the correspondence to the transposed
potential. Observing that $\pha i,1.=\5\pha1,i.$, the above result for
the phase $n/\G$ of the field $X_1$ under a general twist, when applied to
the potential $\W$,  now shows that there are two/zero
monomials that transform with a phase $(\det\r_1)^{-1}$ if $n$ is even/odd.
It is also easy to see directly that, for even $n$, the two monomials
$\prod X_{2i}^{a_{2i}-1}$ and $\prod X_{2i-1}^{a_{2i-1}-1}$ have that property.
In particular, both monomials have the same charge $c/6$, and therefore
the charges of the resulting (a,c) states are indeed asymmetric and of the
correct size. It is not clear to me, however, how to distinguish between
these two monomials, as they have the same transformation properties under all
symmetries. We only know that, for even $n$, the two states that they generate
in the sector $j$ must be the mirror partners of the two states
$\prod \X_{2i}^{a_{2i-1}-1}|0\>$ and $\prod \X_{2i+1}^{a_{2i}-1}|0\>$.%
\footnote{Consideration of selection rules in appropriate orbifolds of the
   mirror pair might settle this question.}

As a by-product of our results we now see that the (c,c) ring is indeed empty,
except for the identity, which is the only invariant monomial.
It only remains to check that the charges of all states are correct, not only
up to an integer. But this follows like before from the inequality
$-1\le (\sum_j \A_j\pha i,j.-q_i)\le0$.
It is also easy to check that, up to normalization of the twist fields,
the orbifold selection rules are concistent with the ring relations
$\6\W/\X_i=\X_{i+1}^{a_i}+a_{i-1}\X_{i-1}\X_i^{a_{i-1}-1}=0$.
The presense of the quantum symmetry, which is isomorphic to $\cg$ and acts
with the correct phases, then implies that we indeed have constructed the
mirror of the chiral ring at the correct point in moduli space.

\section{Chain potentials }

We now use our experience to find the mirror map for
the second type of invertible potential,
\beq W_{chain}=X_1^{a_1}X_2+\ldots+X_{n-1}^{a_{n-1}}X_n+X_n^{a_n}. \eeq
% which forms a chain of fields rather than a closed loop.
Defining $q_{n+1}=0$ we have $a_iq_i+q_{i+1}=1$, with the solution
\beq q_i=\sum_{j=i}^n{\sig{j-i}\0a_i\ldots a_j},~~~~\G=|\cg|=a_1\ldots
a_n.\eeq
This time $j$ generates the whole group $\cg$, but it is again useful
to define a dependent set of group elements which have smaller phases and whose
determinants are related to the weights of the transposed potential.
So we define the transformations $\r_lX_i=\exp(2\p i\pha i,l.)$ with
\beq \pha i,l.={\sig{l-i+1}\0a_i\ldots a_l} ~~~ \hbox{ for }1\le i\le l \eeq
and $\pha l+1,l.=\ldots=\pha n,l.=0$.
Of course only $\r_n$ generates the complete group $\cg$, and we find
\beq (\r_l)^{a_l}\r_{l-1}=1, ~~~~~ \5q_l\eq-\sum_{j=1}^l\pha j,l.=
     -{\ln\det\r_l\02\p i}. \eql{rel}
As the quantities $\5q_i$, defined in this way, obey $a_i\5q_i+\5q_{i-1}=1$
with $\5q_0=0$, they coincide with the charges corresponding to the
transposed potential
\beq \W=\X_1^{a_1}+\X_1\X_2^{a_2}+\ldots+\X_{n-1}\X_n^{a_n}.\eeq
Furthermore, $q_j=-\sum_{i=j}^n\pha j,i.$, implying that
$j^{-1}$ is again the product of all $\r_i$. The twisted ground state
$\ac j^{-2}>$ is the unique state with $Q_R=c/3=-Q_L$.

{}From the generating relations $\6W/\6X_i=0$ for the chiral ring it is easy
to see that $X_n^{a_n}=X_i^{a_i}X_{i+1}=0$ and that we may choose $\a_i<a_i$
for all monomials.
Because of the relations (\ref{rel}) between the transformations $\r_l$,
whose orders are  $a_1\ldots a_l$, we may
represent any twist $\r\aus\cg$ uniquely in the form
$\r=\prod\r_l^{\a_l}$ with $0\le\a_l<a_l$. A simple
calculation shows that the corresponding
phases $\ph_i=\sum_{l=i}^n\a_l\pha i,l.$ obey $-1\le(\ph_i-q_i)\le0$,
where the upper/lower bound is reached iff $n-i$ is odd/even.

So far the situation is very similar to the previous case.
For the present type of potentials, however, the  dimension of the chiral ring
is $|\cR|=%\prod_{i=1}^n{1-q_i\0q_i}=
(1-q_1)\G$.  %=\sum_{i=0}^n\sig i a_{i+1}\ldots a_n \eeq
This number is in general different from
\beq |\5\cR| =(1-\5q_n)\G=\sum_{i=0}^n\sig i a_{1}\ldots a_{n-i}, \eeq
which is the number of states that we should identify in the orbifold $W/\cg$.
In contrast to the previous case, these dimensions are always smaller than
the order $\G$ of the  phase symmetry. Several of the allowed phases thus
cannot be realized in the chiral ring, and, in turn, several twisted
sectors will not contribute any invariant state.
A possible basis for the chiral ring derived from $W_{chain}$ has been
described recursively in ref.~\cite{alg}. It consists of all monomials
$\prod X_i^{\a_i}$ with $\a_1\le a_1-2$ and $\a_i\le a_i-1$ for $2\le i\le n$
or with $\a_1=a_1-1$, $\a_2=0$ and all other exponents fulfilling the same
relation for the potential $W$ with the first two fields set to~0. From
the analysis of the mirror map we will now recover the same basis,
which, as we will see, is the unique choice with $\a_i<a_i$.

Because of the formula for the determinant of $\r_i$ it is not difficult
to guess that the correct mirror map should send $\prod\X^{\A_i}$ into the
sector twisted by $\prod \r_i^{\A_i}$.
This covers the group $\cg$ exactly once if let $0\le\A_i< a_i$.
%We therefore only have to check that the correct sectors survive and that they
%contain one and only one element of the (a,c) ring that has the correct
%%charges
We therefore only have to check that exactly one element of the (a,c) ring
survives in the correct sectors, and that it has the correct charges.

The chain-like structure of our potential implies that $X_{i+1}$ is invariant
under a twist whenever $X_i$ is invariant.
Let, therefore, $\{X_{s},\ldots,X_n\}$ be the sets of fields
invariant under $h'=h/j$. $X_s$ transforms with a phase
\beq \ph_s'=-{\A_s+1\0a_s}+{\A_{s+1}+1\0a_sa_{s+1}}-\ldots-
     {\sig{n-s}\A_n+1\0a_s\ldots a_n},\eeq
which should be integer. Together with $0\le\A_j<a_j$ this implies that
$\A_{n-2i}=a_{n-2i}-1$ and $\A_{n-2i+1}=0$ for all $\A_j$ with $j\ge s$, and
that $\A_{j-1}$ must not follow that pattern.
So we have to consider states of the form
\beq X_s^{\a_s}\ldots X_{n}^{\a_{n}}|\r_1^{\A_1}\ldots\r_{s-1}^{\A_{s-1}}
     \ldots\r_{n-2}^{a_{n-2}-1}\r_{n}^{a_{n}-1}\>.  \eeq
The determinant of the action of $g=\r_n$ on the invariant fields is given by
\beq {1\02\p i}\ln\detat{g},h'. =
     -{1\0a_n}+{1\0a_na_{n-1}}-\ldots-{\sig{n-s}\0a_s\ldots a_{n}},\eeq
which must be compensated modulo 1
by the phase $\sum_{j\ge s} \a_j\pha j,n.$ of the
monomial under the transfomation $g$. As before, we conclude that
$\a_{s+2i}=a_{s+2i}-1$ and $\a_{s+2i+1}=0$ for $i\ge0$. If, however, $n-s$
is even, then the monomial $X_s^{\a_s-1}X_{s+2}^{\a_{s+2}-1}\ldots
X_n^{\a_n-1}$ vanishes in the effective Landau--Ginzburg theory with
potential $W_{|_{h'}}=X_s^{a_s}X_{s+1}+\ldots+X_n^{a_n}$.

We conclude that the projected (a,c) ring consists of all states of the form
\beq X_s^{a_s-1}X_{s+2}^{a_{s+2}-1}\ldots X_{n-1}^{a_{n-1}-1}
     |\r_1^{\A_1}\ldots\r_{s-1}^{\A_{s-1}}~\r_{s+1}^{a_{s+1}-1}
     \r_{s+3}^{a_{s+3}-1}\ldots\r_{n}^{a_{n}-1}\>,             \eeq
with $n-s$ odd and $0\le\A_j<a_j-\d_{j,s-1}$ for $1\le j<s$.
Their charges are indeed anti-symmetric, because $X_i^{a_i}X_{i+1}$ is
homogeneous of degree 1, so that the charge of $X_i^{a_i-1}$ is
$(a_i-1)q_i=1-q_i-q_{i+1}$, in agreement with the criterion given in the
paragraph of eq.~(\ref{qright}).
To check that the value of the right charge is correct, we observe that
the contribution $\sum_{inv}(\2-q_i)$ of the monomial exactly matches the
`missing' contribution $\sum_{inv}(1-q_i-\th_i^{h'})$ in eq.~(\ref{qright}),
because the value $\th_i^{h'}$ for an untwisted fields $X_i$ is $-1$ if $n-i$
is even and $0$ if $n-i$ odd. This implies that the eigenvalue of $\5J_0$ is
$\sum\A_i\5q_i$ for all invariant states, regardless of the number of
untwisted fields.

For the surviving set of exponents we obtain the following description:
Choose some number $1\le s\le n+1$ with $n-s$ odd. Then let
$0\le\A_{j}\le a_{j}-1$ for $j\le s-2$,
$\A_{s+2i}=0$ and $\A_{s+2i+1}=a_{s+2i+1}-1$ for $i\ge0$,
and $0\le\A_{s-1}\le a_{s-1}-2$.
The corresponding basis is indeed identical to the one given in \cite{alg}.
It is also straightforward to check that the defining relations of the chiral
ring are consistent with the twist selection rules, which completes our
proof for the second type of polynomials.

\section{Summary and discussion}

We gave a proof for the BH construction of the mirror by explicitly
constructing the mirror map for invertible potentials. Technically, the
transposition can be traced to the fact that the twist
group can be generated in terms of symmetry transformations $\r_i$, the
determinants of which are related to the weights of the transposed potential
by $\5q_i={i\02\p}\ln\det\r_i$. Consistency of the operator products with
the twist selection rules then implies that the mirror map sends a
monomial $\prod \5X_i^{\A_i}$ into a sector twisted by $\prod \r_i^{\A_i}$.

A necessary condition for the construction to work is that the projection
makes the (c,c) ring trivial. This, in particular, implies that all moduli
are fixed by discrete symmetries. The same should be true for the mirror
model, so that the presense of the quantum symmetry, together with the
matching of the charge degeneracies and the selection rules, provide strong
evidence that the conformal field theories are indeed isomorphic. As a
simple consequence, the BH construction can then be extended to more general
orbifolds.

A remarkable feature of our models is that, except for loops with an even
number of fields, not all representations
of the twist group are present in the chiral ring. In turn, some of the
twisted sectors do not contain any invariant anti-chiral states. For the
remaining sectors, however, we have an identification between twist fields
and certain monomials in the dual theory with flipped left charge.
Using the information on operator
products that is encoded in the dual ring relations,
it should be possible to extract non-trivial information on Yukawa couplings,
thereby extending the results of ref. \cite{cou}.

{\it Acknowledgements.} It is a pleasure to thank Per Berglund,
     Wolfgang Lerche and Harald Skarke for discussions,
     and Martin Winder for his assistance in computations.
% \newpage      % 10baseT or AUI connector (vs. 10base2 coaxial connector)

\end{document}